\documentclass[journal,onecolumn,10pt]{IEEEtran}
\IEEEoverridecommandlockouts
\usepackage{amsmath,amssymb,amsfonts}
\usepackage{algorithm}
\usepackage[noend]{algpseudocode}
\usepackage{graphicx}
\usepackage{setspace}
\usepackage{booktabs}
\usepackage{cite}
\usepackage{xcolor}

\usepackage{enumitem} 
\usepackage{siunitx}
\usepackage{mathtools}
\usepackage[font=footnotesize, labelsep=period, justification=centering]{caption}
\usepackage{dblfloatfix}
\newtheorem{proposition}{Proposition}
\begin{document}
\setstretch{1.5}

\title{Hierarchical Traffic Engineering for Scalable Software-Defined Satellite
Networks }

\author{
\IEEEauthorblockN{Mohamed Khiari\IEEEauthorrefmark{1},
Wiem Abderrahim\IEEEauthorrefmark{2},
Olfa Ben Yahia\IEEEauthorrefmark{1},\\
Fatma Abdelkefi\IEEEauthorrefmark{2},
Wessam Ajib\IEEEauthorrefmark{3},
and Gunes Karabulut Kurt\IEEEauthorrefmark{1}\\}
 \IEEEauthorblockA{\IEEEauthorrefmark{1}Poly-Grames Research Centre, Electrical Engineering Depart., Polytechnique Montréal, Montréal, QC, Canada\\}
 \IEEEauthorblockA{\IEEEauthorrefmark{2}MEDIATRON Lab, Higher School of Communications of Tunis (Sup'Com), University of Carthage, Tunisia\\}
  \IEEEauthorblockA{\IEEEauthorrefmark{3} Computer Sciences Depart. Univ. Québec à Montréal, Montreal, QC, Canada\\}
\IEEEauthorblockA{Emails: mohamed.khiari@etud.polymtl.ca, \{wiem.abderrahim, fatma.abdelkefi\}@supcom.tn, \\\{olfa.ben-yahia, gunes.kurt\}@polymtl.ca, ajib.wessam@uqam.ca}
}

\maketitle

% =========================================================================
\begin{abstract}
%-----------------------------------
Future satellite systems are shifting toward regenerative payloads that interconnect many modem banks into a
reconfigurable on-board network. However, conventional constellation-level routing overlooks intra-payload congestion and load imbalance, which can limit the traffic each satellite can accommodate. This paper proposes a hierarchical software-defined networking (SDN) framework for traffic engineering (TE) across both the constellation and payload levels. The system is represented by a two-level graph, where each satellite node in the constellation graph is expanded into an internal network of interconnected modem banks.  
The proposed hierarchical payload-aware allocation (HPAA) algorithm coordinates the two levels through the exchange of assigned traffic demand and advertised capacity budgets.  
 %TE across both levels is then posed as a bilevel optimization problem that maximizes the minimum served fraction of traffic demands while balancing the internal load of each payload.
% \textcolor{blue}{Simulation results show that HPAA  increases the delivered throughput
% and reduces packet loss compared with payload-blind and shortest-path routing. These findings highlight that hierarchical control accounting for the on-board network constraints of regenerative payloads is a key lever to improve the performance of satellite constellations.} 
{Simulation results show that, under heavy
load, HPAA %increases the delivered throughput by $41\%$ and reduces the end-to-end delay by $22\%$ compared with payload-blind control, and 
delivers about $9$ times the throughput of shortest-path routing, with $17\%$ lower end-to-end delay and no packet loss. These findings highlight that
hierarchical control accounting for the on-board network constraints of
regenerative payloads is a key lever to improve the performance of
satellite constellations.}
\end{abstract}


% =========================================================================
\section{Introduction}\label{sec:intro}
The rapid evolution of low-cost small satellites has led to the
rise of mega-constellation networks in low Earth orbit (LEO),
which provide low-latency broadband communication with
global coverage~\cite{deng2026backpressure}. Through on-board processing (OBP) and
inter-satellite links (ISLs), these systems enhance connectivity and
autonomy \cite{benyahia2025evolution}. However, the increased satellite density and the constant motion
of the constellation increase the routing complexity and require continuous
updates of routing decisions~\cite{kodheli2020survey, 11738729}. Software-defined networking (SDN) is a natural response to this challenge, especially given the industry’s shift from hardware to software~\cite{gao2025regional,11738729}. By separating control-plane functions from data-plane
forwarding and centralizing network management in a programmable
controller, SDN facilitates efficient resource allocation and
flexible service deployment in satellite networks (SatNets)
~\cite{xu2018sdnsat, 11391552}. In large-scale LEO constellations, however, a single controller faces scalability challenges due to limited processing resources and long communication distances between the controller and
satellites. These limitations motivate the adoption of distributed
and hierarchical control-plane architectures
~\cite{papa2018controller,11738729}.

Hierarchical control has been investigated in the literature through three approaches. The first organizes control across orbital
tiers, where LEO, medium Earth orbit (MEO), and geostationary Earth orbit (GEO) layers exchange summarized topology
information~\cite{akyildiz2002mlsr, mao2024fso,11391552}. The second operates
across geographic regions, where the constellation is partitioned into
clusters for scalable routing~\cite{gao2025regional, 11738729}. The third
distributes control across controller layers, where multiple SDN
controllers share a single constellation graph~\cite{jiang2023sdnsurvey,9726849,papa2018controller}.
In all these approaches, the hierarchy is established {between}
satellites such that each satellite is modeled as a single node ~\cite{xu2018sdnsat,papa2018controller,akyildiz2002mlsr, mao2024fso,11391552,gao2025regional, 11738729,jiang2023sdnsurvey,9726849}. A regenerative payload, however, comprises interconnected modem
banks forming an internal forwarding network, where congestion
and load imbalance can limit the satellite's achievable
throughput~\cite{benyahia2025wcl}. Existing constellation-level resource allocation schemes and
SDN control policies generally overlook this intra-satellite
forwarding layer. This limitation motivates a hierarchical SDN
architecture that extends network control below the satellite-node
level. In this architecture, an SDN macro-controller manages
constellation-level traffic allocation, while an onboard SDN
micro-controller manages routing and resource allocation among
the interconnected modem banks of each satellite.
\begin{figure*}[t]
  \centering  
  \includegraphics[height=0.25\textheight,keepaspectratio]{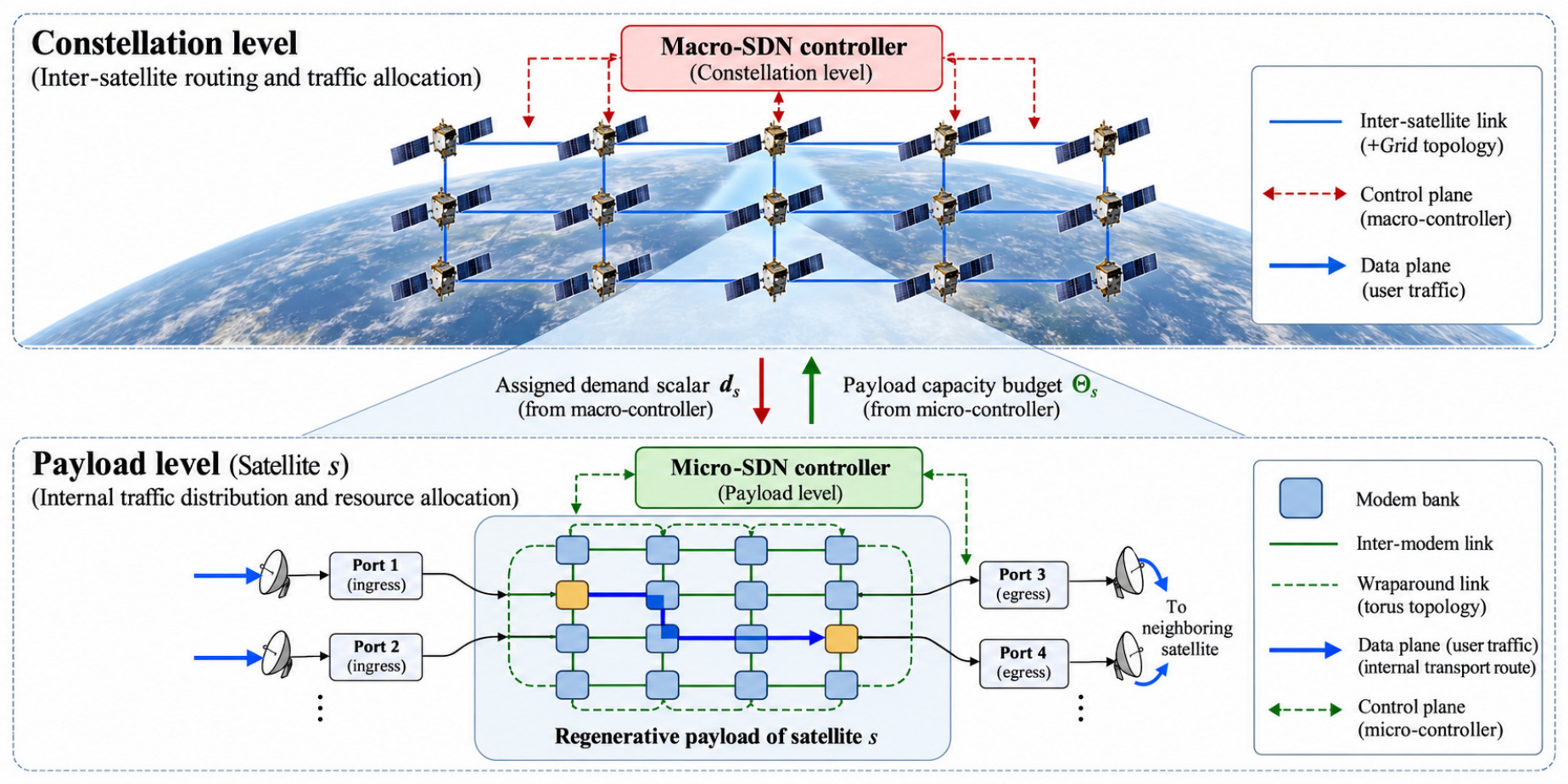}%
\caption{Hierarchical SDN architecture for joint constellation-level traffic allocation and intra-payload load balancing.}
\label{fig:model}
\end{figure*}

Therefore, this paper proposes a hierarchical SDN-based traffic
engineering (TE) framework for Satellite Networks (SatNets), building
upon the programmable payload architecture introduced
in~\cite{benyahia2025evolution,benyahia2025wcl}. We introduce a hierarchical graph model in which each satellite
node in the constellation graph is expanded into a payload graph
of interconnected modem banks, capturing intra-payload congestion
and load imbalance.
The hierarchical TE problem is formulated as a bilevel
optimization problem. At the constellation level, the objective
is to maximize the minimum fraction of served traffic demand,
while at the payload level, the objective is to balance the
load across modem banks. The two levels are coupled through the traffic demand assigned
to each satellite and the capacity budget advertised by its
payload. We show that, under the demultiplexer model, this advertised
budget per satellite is sufficient to achieve the same optimal served
fraction as an ideal centralized controller. Simulation results demonstrate that the proposed hierarchical payload-aware allocation (HPAA) algorithm outperforms baseline
control and routing strategies in terms of delivered throughput,
packet loss, and end-to-end delay. 

The remainder of this paper is organized as follows.
Section~\ref{sec:model} introduces the system model,
while Section~\ref{sec:formulation} presents the optimization
formulations. Section~\ref{sec:solution} describes the proposed
solution approach, followed by the numerical results in
Section~\ref{sec:sim}.
Finally, Section~\ref{sec:conclusion} concludes the paper.

%=====================================================

% =========================================================================

\section{System Model}\label{sec:model}
Our system model comprises a set 
of ground stations that exchange traffic through a constellation of 
satellites, as depicted in Fig.~\ref{fig:model}. Each satellite carries a regenerative payload composed of 
interconnected modem banks arranged in a torus topology. The satellites are interconnected through time-varying ISLs.  To provide path
diversity, each ground station is assumed to maintain simultaneous
links with several satellites. 
\textit{Notation:} Calligraphic letters (e.g., $\mathcal{S}$) denote
sets, $|\mathcal{S}|$ denotes the cardinality of set $\mathcal{S}$ and bold lowercase letters (e.g., $\mathbf{x}$) denote vectors.

\subsection{Constellation-Level Model}
Time is divided into snapshots indexed by
$t \in \{0, 1, 2, \dots\}$. The topology is fixed within a snapshot. 
Let $\mathcal{S}$ denote the set of satellites and $\mathcal{G}$ the
set of ground stations. The constellation at the snapshot $t$ is modeled as
a directed graph
\begin{equation}
G^{(1)}(t) = \big(\mathcal{V},\, \mathcal{E}(t)\big), \qquad
\mathcal{V} = \mathcal{S} \cup \mathcal{G},
\end{equation}
where $\mathcal{E}(t)$ contains a directed edge for each ISL and each
ground--satellite link available at snapshot $t$, derived from
ephemeris data. The graph is directed to capture the asymmetry of link
capacities between the two directions of a link, such as between feeder
uplinks and downlinks. Each edge $e \in \mathcal{E}(t)$ has a capacity
$C_e$. 

\subsection{Payload-Level Model}
Each satellite $s \in \mathcal{S}$ carries a regenerative payload
composed of $M$ interconnected modem banks,
modeled as a directed graph $G^{(2)}_s = (\mathcal{V}_s,\, \mathcal{E}_s),$ 
whose vertices are the modem banks and whose edges are the inter-modem
links, arranged in a toroidal topology, following the payload
 architecture of~\cite{benyahia2025wcl}. 
%Directedness reflects the
% physical traversal of the payload, from the demultiplexer (ingress)
% stage to the multiplexer (egress) stage. In contrast to the
% constellation graph, the payload graph is time-invariant: the
% intra-payload interconnect is wired, and neither its vertex set nor its
% edge set varies across snapshots. 
Each modem bank $v \in \mathcal{V}_s$ has a service rate $\mu_v$
bounding the total traffic it can process. 
An intra-payload edge $(u,v) \in \mathcal{E}_s$ represents the unidirectional
inter-modem link from modem bank $u$ to modem bank $v$, with
$u, v \in \mathcal{V}_s$.
Each intra-payload edge
$(u,v) \in \mathcal{E}_s$ has a capacity $c_{uv} > 0$. Each payload is characterized, at every snapshot, by a scalar
\emph{payload budget} $\Theta_s \ge 0$, defined as the maximum total
traffic rate that the payload of satellite $s$ can process given its
link capacities and service rates. The budget is advertised by
satellite $s$ at each snapshot to the
constellation level. 
Each payload is attached to the constellation through a demultiplexer
and a multiplexer, connected to modem banks referred to as
\emph{ingress} and \emph{egress} ports, respectively. Under this
\emph{demultiplexer model}, traffic arriving on any link can be
distributed over all ingress ports, and traffic leaving on any link
can be collected from all egress ports.

\subsection{Traffic Model}
Traffic consists of a set $\mathcal{K}$ of end-to-end traffic
demands. Each demand $k \in \mathcal{K}$ is a triple
$(\sigma^{(k)}, \delta^{(k)}, d^{(k)})$, where $\sigma^{(k)} \in \mathcal{G}$ is the
source ground station, $\delta^{(k)} \in \mathcal{G}$ is the destination
ground station, and $d^{(k)} > 0$ is the requested rate. The packets of demand $k$ are generated according to a Poisson process
of rate $d^{(k)}$, independently of the other demands. % such that the times $T^{(k)}$ between consecutive packets are independent and exponentially distributed,
% \begin{equation}\label{eq:poisson}
% \Pr\big\{T^{(k)} > t\big\} \;=\; e^{-d^{(k)} t}, \qquad t \ge 0.
% \end{equation}
For each demand $k$, a set $\mathcal{P}^{(k)}$ of candidate paths in $G^{(1)}(t)$
from $\sigma^{(k)}$ to $\delta^{(k)}$ is computed at each snapshot as the
$K_{\mathrm{p}}$ shortest paths, where $K_{\mathrm{p}}$ is a fixed
integer parameter. The traffic of a demand may be split across its candidate paths.

\section{Problem Formulation }\label{sec:formulation}

In this section, we first formulate the TE problem
over the two-level model. Then, we formulate the same
problem for an ideal centralized controller and show that both achieve
the same optimal served fraction.

\subsection{Bilevel Problem Formulation}
\subsubsection{Constellation-level problem}\label{sec:P1}
The goal of the constellation-level problem is to share the network capacity fairly among the traffic demands. Let $x^{(k)}_P \in \mathbb{R}_{+}$ denote the rate of demand $k$ carried on
path $P \in \mathcal{P}^{(k)}$. The rates collected
in the vector $\mathbf{x}$   are
allocated to prevent the starvation of
demands that traverse congested regions of the constellation. Accordingly, we express the objective function through the
max--min fairness criterion applied to the {served fractions},  as
follows:
\begin{equation}\label{eq:objective}
\max_{\mathbf{x}}\;\; \min_{k \in \mathcal{K}}\; f^{(k)}(\mathbf{x}) ,
\end{equation}
where the served fraction of demand $k$ under an allocation
$\mathbf{x}$ is given by:
\begin{equation}\label{eq:servedfrac}
f^{(k)}(\mathbf{x}) \;=\; \frac{1}{d^{(k)}}\sum_{P\in\mathcal{P}^{(k)}} x^{(k)}_P ,
\end{equation}
in which $\displaystyle\sum_{P} x^{(k)}_P$, the total rate granted to demand $k$
across its candidate paths, is normalized by the requested rate $d^{(k)}$
to ensure fairness between customers with heterogeneous demands.
Feasible allocations should meet specific constraints, grouped into
traffic requirements and resource capacities.

\textit{a) Traffic Requirement Constraints:}
The rate granted to a commodity cannot exceed its requested rate:
\begin{equation}\label{eq:P1dem}
\text{C}_1 :\quad \sum_{P\in\mathcal{P}^{(k)}} x^{(k)}_P \;\le\; d^{(k)},
\qquad \forall k\in\mathcal{K}.
\end{equation}

\textit{b) Resource Capacity Constraints:}
The allocation is also limited by the transmission and processing
resources of the network. For an edge $e \in \mathcal{E}(t)$, let
$\mathcal{P}^{(k)}_e \subseteq \mathcal{P}^{(k)}$ denote the subset of
candidate paths of demand $k$ that traverse $e$; similarly,
$\mathcal{P}^{(k)}_s$ denotes the subset traversing satellite
$s \in \mathcal{S}$. On the one hand, the total flow traversing any
constellation-level edge cannot exceed its capacity:
\begin{equation}\label{eq:P1link}
\text{C}_2 :\quad
\sum_{k\in\mathcal{K}}\;\sum_{P\in\mathcal{P}^{(k)}_e} x^{(k)}_P
\;\le\; C_e,
\qquad \forall e\in\mathcal{E}(t).
\end{equation}
On the other hand, the total flow traversing any satellite cannot
exceed its advertised payload budget:
\begin{equation}\label{eq:P1budget}
\text{C}_3 :\quad
\sum_{k\in\mathcal{K}}\;\sum_{P\in\mathcal{P}^{(k)}_s} x^{(k)}_P
\;\le\; \Theta_s,
\qquad \forall s\in\mathcal{S}.
\end{equation}

Combining the objective~\eqref{eq:objective} with the  constraints~\eqref{eq:P1dem},~\eqref{eq:P1link} and~\eqref{eq:P1budget}
yields the constellation-level problem:
\begin{align}
\textbf{(P1)} :\qquad
&\max_{\mathbf{x} \ge \mathbf{0}}\;\;
\min_{k \in \mathcal{K}}\; f^{(k)}(\mathbf{x}) \label{eq:P1}\\
&\;\text{s.t.}\quad \text{C}_1 - \text{C}_3. \nonumber
\end{align}
% In the sequel, $\mathcal{X}$ denotes the feasible set of (P1), i.e.,
% the set of vectors $\mathbf{x} \ge \mathbf{0}$
% satisfying~\eqref{eq:P1dem}--\eqref{eq:P1budget}.

%%%%% without the demultiplexer model 
%Given a solution $\mathbf{x}^\star$ of (P1), the total rate entering
% the payload of satellite $s$ at port $i$ and exiting at port $j$ is
% given by:
% \begin{equation}\label{eq:aggregation}
% d^{\,ij}_s \;=\; \sum_{k\in\mathcal{K}}\;
% \sum_{P\in\mathcal{P}^{(k)}_s(i,j)} x^{k\star}_P ,
% \end{equation}
% where $\mathcal{P}^{(k)}_s(i,j) \subseteq \mathcal{P}^{(k)}_s$ denotes the
% subset of candidate paths of demand $k$ that enter the payload on a
% link assigned by $\pi_s$ to port $i$ and leave it on a link assigned
% to port $j$. These rates are collected in the \emph{port-to-port
% demand matrix} $D_s = \big[d^{\,ij}_s\big]_{i,j \in \mathcal{B}_s}$;
% the matrices $\{D_s\}_{s\in\mathcal{S}}$ constitute the assigned
% demands of the payload-level problem.
% \vspace{5pt}

Given a solution $\mathbf{x}^\star$ of \textbf{(P1)}, the total rate
routed through the satellite $s$ is given by:
\begin{equation}\label{eq:aggregation}
d_s \;=\; \sum_{k\in\mathcal{K}}\;
\sum_{P\in\mathcal{P}^{(k)}_s} x^{k\star}_P ,
\end{equation}
where $\mathcal{P}^{(k)}_s \subseteq \mathcal{P}^{(k)}$ denotes the subset of
candidate paths of demand $k$ that traverse satellite $s$. Under the
demultiplexer model, $d_s$ constitutes the demand assigned to the
payload of satellite $s$.
\vspace{5pt}

%%%\subsubsection{Payload-Level
%%%% without the demultiplexer model
% Let $\mathcal{K}_s$
% denote the set of \emph{active port pairs} of satellite $s$,
% \begin{equation}\label{eq:portpairs}
% \mathcal{K}_s \;=\;
% \big\{\, k_s = (i,j) \in \mathcal{B}_s \times \mathcal{B}_s
% \;\big|\; d^{\,ij}_s > 0 \,\big\},
% \end{equation}
%  For each
% pair $k_s = (i,j) \in \mathcal{K}_s$, let $\mathcal{Q}^{k_s}$ denote
% the set of candidate paths from port $i$ to port $j$ in $G^{(2)}_s$.

% Let $y^{k_s}_Q \in \mathbb{R}_{+}$ denote the
% rate of the assigned demand $d^{\,ij}_s$ carried on path
% $Q \in \mathcal{Q}^{k_s}$. The rates collected in $\mathbf{y}$ are chosen such that the most
% loaded inter-modem link retains the largest possible spare capacity.  Therefore, we express the objective function
% through the max--min criterion applied to the residual capacities of
% the inter-modem links as follows:
% \begin{equation}\label{eq:P2objective}
% \max_{\mathbf{y}}\;\; \min_{(u,v)\in \mathcal{E}_s}\;
% \varphi_{uv}(\mathbf{y}) ,
% \end{equation}
\subsubsection{Payload-level problem}\label{sec:P2}
The goal of the payload-level problem is to balance the traffic load across the inter-modem links. Let $\mathcal{Q}_s$ denote the set of
candidate paths in $G^{(2)}_s$ from an ingress port to an egress port,
and let $y_Q \in \mathbb{R}_{+}$ denote the rate carried on path
$Q \in \mathcal{Q}_s$. The rates collected in $\mathbf{y}$ are chosen
such that the most loaded inter-modem link retains the largest
possible spare capacity. Therefore, we express the objective function
through the max--min criterion applied to the residual capacities of
the inter-modem links as follows:
\begin{equation}\label{eq:P2objective}
\max_{\mathbf{y}}\;\; \min_{(u,v)\in \mathcal{E}_s}\;
\varphi_{uv}(\mathbf{y}) ,
\end{equation}
%%%% without demultiplexer model
% where the residual capacity of edge $(u,v)$ under an assignment
% $\mathbf{y}$ is given by:
% \begin{equation}\label{eq:phires}
% \varphi_{uv}(\mathbf{y}) \;=\; c_{uv} \;-\;
% \sum_{k_s\in\mathcal{K}_s}\;\sum_{Q\in\mathcal{Q}^{k_s}_{uv}}
% y^{k_s}_Q ,
% \end{equation}
% in which $\mathcal{Q}^{k_s}_{uv} \subseteq \mathcal{Q}^{k_s}$ denotes
% the subset of candidate paths of pair $k_s$ that traverse $(u,v)$, so
% that the double summation is the total flow routed through $(u,v)$.
% Feasible assignments should meet specific constraints, grouped into
% traffic requirements and resource capacities.
where the residual capacity of an edge $(u,v)$ under an assignment
$\mathbf{y}$ is given by:
\begin{equation}\label{eq:phires}
\varphi_{uv}(\mathbf{y}) \;=\; c_{uv} \;-\;
\sum_{Q\in\mathcal{Q}_{s,uv}} y_Q ,
\end{equation}
in which $\mathcal{Q}_{s,uv} \subseteq \mathcal{Q}_s$ denotes the
subset of candidate paths that traverse $(u,v)$, where $\displaystyle\sum_{Q\in\mathcal{Q}_{s,uv}}\!\! y_Q$
is the total rate routed through $(u,v)$. 
Feasible assignments should
meet specific constraints, grouped into traffic requirements and
resource capacities.

\textit{a) Traffic requirement constraints:}
Each assigned demand is served exactly: 
\begin{equation}\label{eq:P2dem}
\text{C}_4 :\quad \sum_{Q\in\mathcal{Q}_s}\!\!  y_Q \;=\; d_s .
\end{equation}

\textit{b) Resource capacity constraints:}
The total flow routed through any inter-modem link cannot exceed its
capacity:
\begin{equation}\label{eq:P2cap}
\text{C}_5 :\quad \sum_{Q\in\mathcal{Q}_{s,uv}}\!\!  y_Q \;\le\; c_{uv},
\qquad \forall (u,v)\in \mathcal{E}_s .
\end{equation}

In addition, the total flow processed by a modem bank cannot exceed
its service rate. Denoting by $\mathcal{Q}_{s,v} \subseteq \mathcal{Q}_s$
the subset of candidate paths that traverse bank $v$, this constraint
is written as follows:
\begin{equation}\label{eq:P2mu}
\text{C}_6 :\quad \sum_{Q\in\mathcal{Q}_{s,v}}\!\!  y_Q \;\le\; \mu_v,
\qquad \forall v\in \mathcal{V}_s .
\end{equation}
Combining the objective~\eqref{eq:P2objective} with the  constraints~\eqref{eq:P2dem},~\eqref{eq:P2cap} and~\eqref{eq:P2mu} yields the payload-level problem of satellite $s$:
\begin{align}
\textbf{(P2)} :\qquad
&\max_{\mathbf{y} \ge \mathbf{0}}\;\;
\min_{(u,v)\in \mathcal{E}_s}\; \varphi_{uv}(\mathbf{y}).
\label{eq:P2}\\
&\;\text{s.t.}\quad \text{C}_4 - \text{C}_6. \nonumber
\end{align}
In turn, the payload-level problem determines the budget $\Theta_s$
used in constraint~C$_3$ of problem \textbf{(P1)}. It is defined as the maximum total flow that
the payload of satellite $s$ can carry under its resource capacity
constraints $\text{C}_5$ and $ \text{C}_6$ as follows:
\begin{equation}\label{eq:budget}
\Theta_s \;=\;
\max_{\mathbf{y} \ge \mathbf{0}}\;
\Big\{\, \sum_{Q\in\mathcal{Q}_s} y_Q \;\text{; such that}\; \text{C}_5 \text{ and }
\text{C}_6 \text{ hold} \,\Big\}.
\end{equation}
The budgets
$\{\Theta_s\}_{s\in\mathcal{S}}$ appearing in constraint~C$_3$ of
\textbf{(P1)} are determined at the payload level
through~\eqref{eq:budget}, while the assigned demands
$\{d_s\}_{s\in\mathcal{S}}$ defining constraint~C$_4$ of
\textbf{(P2)} are determined by the solution of \textbf{(P1)}
through~\eqref{eq:aggregation}. The joint problem is therefore a bilevel
program, with \textbf{(P1)} as the leader and the per-satellite
problems \textbf{(P2)} as the followers.

\subsection{Centralized
Problem Formulation}

As a reference, an ideal centralized controller with knowledge of every
payload solves the problem on the flat graph obtained from $G^{(1)}$ by
replacing each satellite node $s \in \mathcal{S}$ with its payload graph
$G^{(2)}_s$. The nodes of this graph are the ground stations and the
modem banks of all satellites, and its edges are the ground--satellite
links, the ISLs, and the inter-modem links of all satellites. The centralized problem can be formulated as follows:
\begin{align}
\textbf{(P0)} :\quad
& \max_{\mathbf{x} ,\;
\mathbf{y}_{s\in\mathcal{S}}}\;\;
\min_{k\in\mathcal{K}}\; f^{(k)}(\mathbf{x}) \label{eq:P0}\\
&\;\text{s.t.}\quad \text{C}_1\; \text{--C}_2, \nonumber\\
&\phantom{\;\text{s.t.}\quad} \text{C}_4\text{--C}_6, \;
\forall s\in\mathcal{S}, \nonumber
\end{align}
where $\mathbf{x}$ is the traffic on the ground--satellite links
and the ISLs and $\mathbf{y}_s$ indicates the traffic on the inter-modem links of
satellite $s$, both chosen jointly. The demand in C$_4$ is given by:
\begin{equation}\label{eq:dsP0}
d_s \;=\; \sum_{k\in\mathcal{K}}\;\sum_{P\in\mathcal{P}^{(k)}_s} x^{(k)}_P.
\end{equation}

Problem \textbf{(P0)} requires knowledge of the
internal state of every payload at the controller. The following result
shows that, under the demultiplexer model, hierarchical control based
only on the budgets achieves the same optimal served
fraction.

\begin{proposition}\label{prop:exact}
Under the demultiplexer model, solving \textbf{(P1)} then
\textbf{(P2)} achieves the same optimal served fraction of \textbf{(P0)}.
\end{proposition}

\begin{IEEEproof}
First, let $\mathbf{x}$ and $\{\mathbf{y}_s\}$ be feasible for
\textbf{(P0)}. At every satellite $s$, $\mathbf{y}_s$ satisfies
C$_4$--C$_6$ with demand $d_s$, so it is a feasible solution of
\textbf{(P2)}. In particular, it satisfies C$_5$ and C$_6$, so it is one
of the assignments over which $\Theta_s$ is maximized
in~\eqref{eq:budget}; since it carries $d_s$, it follows that
$d_s \le \Theta_s$. Hence, $\mathbf{x}$ satisfies C$_3$ and is feasible
for \textbf{(P1)}.

Second, let $\mathbf{x}$ be feasible for \textbf{(P1)}. By C$_3$,
$d_s \le \Theta_s$ at every satellite $s$. Let $\hat{\mathbf{y}}_s$ be
an assignment attaining $\Theta_s$ in~\eqref{eq:budget}, and let
$\mathbf{y}_s = \frac{d_s}{\Theta_s}\,\hat{\mathbf{y}}_s$. Since
$d_s/\Theta_s \le 1$, the rates of $\mathbf{y}_s$ do not exceed those
of $\hat{\mathbf{y}}_s$, so $\mathbf{y}_s$ satisfies C$_5$ and C$_6$;
since it carries exactly $d_s$, it also satisfies C$_4$. Hence,
$\mathbf{y}_s$ is a feasible solution of \textbf{(P2)} at every
satellite. Moreover, $\mathbf{x}$ satisfies C$_1$ and C$_2$, being
feasible for \textbf{(P1)}, so $\mathbf{x}$ and $\{\mathbf{y}_s \in \mathcal{S}\}$ are
feasible for \textbf{(P0)}. Since the bilevel problem and the
centralized problem have the same feasible set of rates $\mathbf{x}$ and both maximize the same function $\min_k f^{(k)}(\mathbf{x})$ over
this set, they achieve the same optimal served fraction.
\end{IEEEproof}

% =========================================================================
\section{Solution Approach}\label{sec:solution}
In this section, we reformulate \textbf{(P1)} and \textbf{(P2)} as
linear programs. Then, we present the proposed algorithm that solves the bilevel problem and analyze its complexity. 

\subsection{Constellation-level linear program}\label{sec:P1prime}
The objective of \textbf{(P1)} is a pointwise minimum, which is not a
linear function of the decision variables. The problem is therefore
recast through its epigraph reformulation, in which the auxiliary
variable $z$ bounds every served fraction
and is maximized over the same feasible set:
\begin{align}
\textbf{(P1$'$)} :\qquad
&\max_{\mathbf{x} \ge \mathbf{0},\; z \ge 0}\;\; z
\label{eq:P1prime}\\
&\;\text{s.t.}\quad \text{C}_1 - \text{C}_3, \nonumber\\
&\phantom{\;\text{s.t.}\quad}
f^{(k)}(\mathbf{x}) \;\ge\; z, \quad \forall k\in\mathcal{K}.
\nonumber
\end{align}
At the optimum, the added constraint is tight for the least served
demand, so $z^\star$ equals the optimal value of \textbf{(P1)}
and the optimal path flows of the two problems coincide.
Problem~\textbf{(P1$'$)} is a linear program, which is solved
efficiently with GLPK through the CVXPY
modeling interface.

\subsection{Payload-level linear program}\label{sec:P2prime}
The same reformulation applies to \textbf{(P2)}, where the auxiliary
variable $\rho_s \ge 0$ bounds the residual capacity of every
inter-modem link:
\begin{align}
\textbf{(P2$'$)} :\qquad
&\max_{\mathbf{y} \ge \mathbf{0},\; z_s \ge 0}\;\; z_s
\label{eq:P2prime}\\
&\;\text{s.t.}\quad \text{C}_4,\; \text{C}_6, \nonumber\\
&\phantom{\;\text{s.t.}\quad}
\varphi_{uv}(\mathbf{y}) \;\ge\; z_s, \quad
\forall (u,v)\in \mathcal{E}_s. \nonumber
\end{align}
At the optimum, $z_s^\star$ equals the residual capacity of the
most loaded inter-modem link, so \textbf{(P2$'$)} and \textbf{(P2)}
share the same optimal assignments. Since $z_s \ge 0$, the added
constraint implies C$_5$, which is therefore omitted.
Problem~\textbf{(P2$'$)} is a linear program, 
{which is solved efficiently with GLPK via the CVXPY
modeling interface.}

\subsection{Proposed algorithm}\label{sec:sol}
Using the results above, we develop Algorithm~\ref{alg:hpaa}, such that each satellite
first computes its budget from the current state of its modem banks. Then,
\textbf{(P1$'$)} is solved by the SDN macro-controller and each
problem \textbf{(P2$'$)} is solved by the SDN micro-controller on board the corresponding
satellite. 

\begin{algorithm}[h]
\caption{Hierarchical Payload-Aware Allocation (HPAA)}
\label{alg:hpaa}
\begin{algorithmic}[1]
\Require $G^{(1)}(t)$, $\{G^{(2)}_s\}_{s\in\mathcal{S}}$, demands
$\mathcal{K}$, paths $\{\mathcal{P}^{k}\}_{k\in\mathcal{K}}$
\ForAll{$s \in \mathcal{S}$ \textbf{in parallel}}
  \State compute the budget $\Theta_s$ from the current state of the
  payload through~\eqref{eq:budget}
\EndFor
\State solve \textbf{(P1$'$)} with the budgets
$\{\Theta_s\}_{s\in\mathcal{S}}$; obtain the rates $\mathbf{x}^\star$
\State compute the assigned demands $\{d_s\}_{s\in\mathcal{S}}$ from
$\mathbf{x}^\star$ through~\eqref{eq:aggregation}
\ForAll{$s \in \mathcal{S}$ \textbf{in parallel}}
  \State solve \textbf{(P2$'$)} with demand $d_s$; obtain the rates
  $\mathbf{y}^\star_s$
\EndFor
\State \textbf{return} the rates $\mathbf{x}^\star$ and
$\{\mathbf{y}^\star_s\}_{s\in\mathcal{S}}$
\end{algorithmic}
\end{algorithm}

Problems \textbf{(P1$'$)} and \textbf{(P2$'$)} are linear programs,
solved to optimality at each snapshot. 
% By its definition
% in~\eqref{eq:budget}, the budget $\Theta_s$ depends only on the link
% capacities and service rates in C$_5$ and C$_6$. Moreover, since the
% traffic entering satellite $s$ can be distributed over all its
% ingress ports and collected from all its egress ports, the budget
% does not depend on the demand assigned to the payload. Once the budgets
% are recomputed from the current states of the modem banks, they no
% longer change, and Algorithm~\ref{alg:hpaa} is therefore guaranteed to converge.
The complexity of each snapshot is determined by \textbf{(P1$'$)},
\textbf{(P2$'$)}, and the budget computation in~\eqref{eq:budget},
which are linear programs solved by interior-point methods in
$\mathcal{O}(n^{3.5})$ operations for $n$ variables and
constraints~\cite{karmarkar1984}. Since each node has a bounded number
of links, the complexity of \textbf{(P1$'$)} is
$\mathcal{O}\big((|\mathcal{K}|K_{\mathrm{p}}+|\mathcal{S}|)^{3.5}\big)$.
The budget computation and \textbf{(P2$'$)} involve the same paths and
constraints, and are solved in parallel at each satellite in
$\mathcal{O}\big((|\mathcal{Q}_s|+M)^{3.5}\big)$. The overall
complexity per snapshot is therefore
$\mathcal{O}\big((|\mathcal{K}|K_{\mathrm{p}}+|\mathcal{S}|)^{3.5}
+ (|\mathcal{Q}_s|+M)^{3.5}\big)$.

\section{Simulations and Results} \label{sec:sim}
In this section, we establish the simulation setup. Then, we evaluate the performance of HPAA. 
\subsection{Simulation Setup} 
% We consider a Low
% Earth Orbit (LEO) constellation composed of 12 satellites across $3$ orbital
% planes.
% Orbit descriptions (TLEs) and high-fidelity position/distance traces are
% generated across $11$ temporal snapshots using the \texttt{Hypatia} LEO
% simulation framework based on the SGP4 orbit propagation model.  A set of 6 ground
% stations is deployed in London, Sydney, New York, Tokyo,
% S\~{a}o Paulo, and Nairobi. 
We consider a lightspeed-sized LEO constellation comprising $198$ satellites, matching the constellation size of Telesat Lightspeed~\cite{liang2025green}. The satellites are distributed across $14$ orbital planes, with $12$ planes containing $14$ satellites and $2$ containing $15$, in a Walker shell at an altitude of $1325$~km and an inclination of $50.88^\circ$~\cite{telesat2020fcc}. %, corresponding to Telesat's inclined sub-constellation~\cite{telesat2020fcc}. 
The satellites are interconnected using a $+Grid$ topology~\cite{bhattacherjee2019topology}.
%, with four optical inter-satellite links (ISLs) per satellite: two intra-plane links and two inter-plane links. This topology is commonly used to model LEO satellite constellations~\cite{bhattacherjee2019topology}. 
%Link propagation delays are computed dynamically from the satellite geometry in \texttt{Hypatia}~\cite{kassing2020hypatia}. %, as the straight-line distance between link endpoints divided by the speed of light. 
%We consider $10$ ground stations (GSs) distributed across the coverage area ~\cite{liang2025green}. %Each GS is connected to its three nearest satellites, subject to a maximum of six links per satellite across ISLs and ground links, corresponding to the six available payload ports.
 The data plane execution
is realized using a custom discrete-event simulator built in
\texttt{SimPy}. %~\cite{simpy}. %Modem banks and all links are  modeled as independent first in first out (FIFO) queues operating with a finite buffer of capacity $B$ packets and a tail-drop policy.
 Table~\ref{tab:const_params} lists
the main simulation settings. We benchmark HPAA against the centralized controller by solving
\textbf{(P0)} besides four baseline control and routing strategies, under
identical simulation settings and hardware parameters:
\begin{itemize}
\item{Payload-blind controller (PBC):} Models a macro-level
SDN controller that ignores internal payload constraints. %$(\mathbf{P1})$ is solved without the payload budget constraint $C_3$.
\item {Single-bank satellite (SBS):} Models
 non-distributed payloads where each satellite is
represented as a single node.% in $G^{(1)}$.

% \item{Load aware-Shortest Path (LSP) :} Traffic demands are split into six equal pieces and routed sequentially using load-aware Dijkstra’s algorithm. Link and modem-bank costs increase with utilization, encouraging traffic to use less-loaded paths. The controller relies solely on current network loads, without enforcing the payload budget \(\Theta_s\) or regulating demand rates.
%\item{Load-Aware Shortest Path (LSP):} Each traffic demand is split into $K_{\mathrm{p}}$ equal pieces, so that it can be spread over as many paths as BCIA has candidates, and the pieces are routed sequentially, one piece of every demand per round,using a load-aware Dijkstra's algorithm.
\item{Load-aware shortest path (LSP):} 
Routes traffic sequentially by using a load-aware Dijkstra's algorithm.
% Each traffic demand is split into $K_{\mathrm{p}}$ equal rates routed sequentially by using a load-aware Dijkstra's algorithm.
\item {Shortest path (SP):} Routes traffic
along the single minimum-hop satellite path using the Dijkstra's algorithm. 
\end{itemize}
\subsection{Results} \label{sec:sim_results}
We evaluate the network performance of HPAA, analyze its sensitivity, and evaluate its
computation time.
\begin{table}[t]
\caption{Simulation settings.}
\label{tab:const_params}
\centering
\footnotesize
\begin{tabular}{|l|l|}
\hline
\textbf{Parameter} & \textbf{Value} \\
\hline
%Number of satellites  & 198 \\
Number of ISL & 2 intra-plane  and 2 inter-plane \\
%Altitude & $1325$ km \\
%Inclination & \ang{50.88} \\
Number of ground stations  & 10 ~\cite{liang2025green}  \\
Number of ground station links & 3  \\
ISL capacity & 5-10~Gbit/s \\
Candidate paths $K_p$ & 24 \\
Number of modem banks $M$  &  16  \\
Number of ingress/egress ports & 6 / 6 border banks \\
Modem service rate  & 100 000 packets/s \\
Inter-modem link $c_{uv}$ & 10~Gbit/s \\
Packet size & 512~bytes \\
Buffer $B$ & $10^3$ packets \\
\hline
\end{tabular}
\end{table}

\begin{figure*}[b]
\centering
\includegraphics[width=1\textwidth]{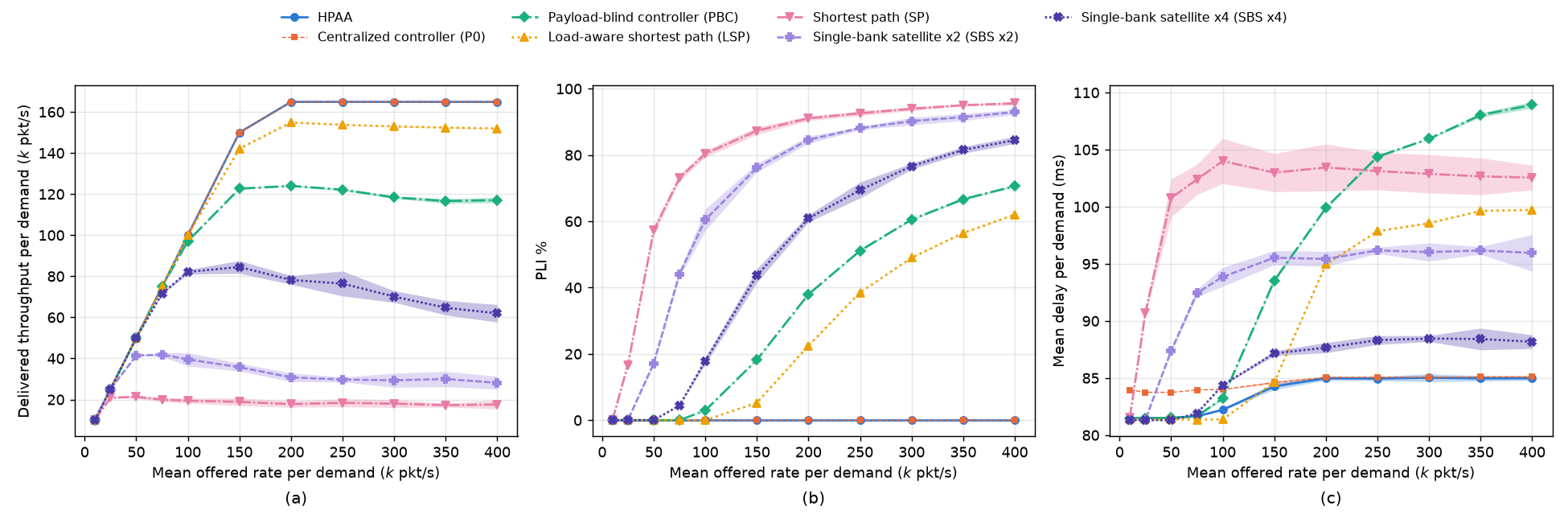}
\caption{ Network performance metrics versus the requested rate per demand, with 95\% confidence intervals.}
\label{fig:sim_results}
\end{figure*}
\subsubsection{Network performance}
We compare HPAA with the  baselines in terms of the delivered throughput,
the packet loss indicator (PLI), and the average end-to-end delay. Performance metrics are
averaged over independent replications, and 95\% confidence intervals
are included. Fig.~\ref{fig:sim_results} shows the performance metrics as a function
of the mean requested rate per demand. Specifically, Fig.~\ref{fig:sim_results}a depicts
the delivered throughput by HPAA, which achieves the highest
throughput of all schemes. HPAA saturates at $165$~kpkt/s as the most loaded satellites reach their
budgets $\Theta_s$. The Centralized Controller (P0) reaches the same
throughput, so the bilevel decomposition loses nothing with respect to
the centralized solution. The baselines follow HPAA at light load but
fall behind once payloads are overloaded.
% saturating at $155$~kpkt/s,
% $124$~kpkt/s and $85$~kpkt/s respectively,
% while the Shortest Path saturates at only $21$~kpkt/s. 
Fig.~\ref{fig:sim_results}b
shows that HPAA, like P0, are fully reliable for
all loads, since they admit only the traffic that the payloads can
switch. In contrast, all baselines show a significant rise in PLI once
their bottleneck modem banks saturate. %For instance, the packet loss reaches $62$\% for LSP and $96$\% for SP at $400$~kpkt/s. 
Fig.~\ref{fig:sim_results}c demonstrates that HPAA has consistently low delays. P0 shows a
higher delay at light load, since among the routings that
reach the same throughput, it selects longer candidate paths. In contrast, the baselines exhibit higher delays, particularly
noticeable at higher traffic demands.

\subsubsection{Sensitivity analysis}
\begin{figure*}[t]
\centering
  \includegraphics[width=0.75\textwidth]{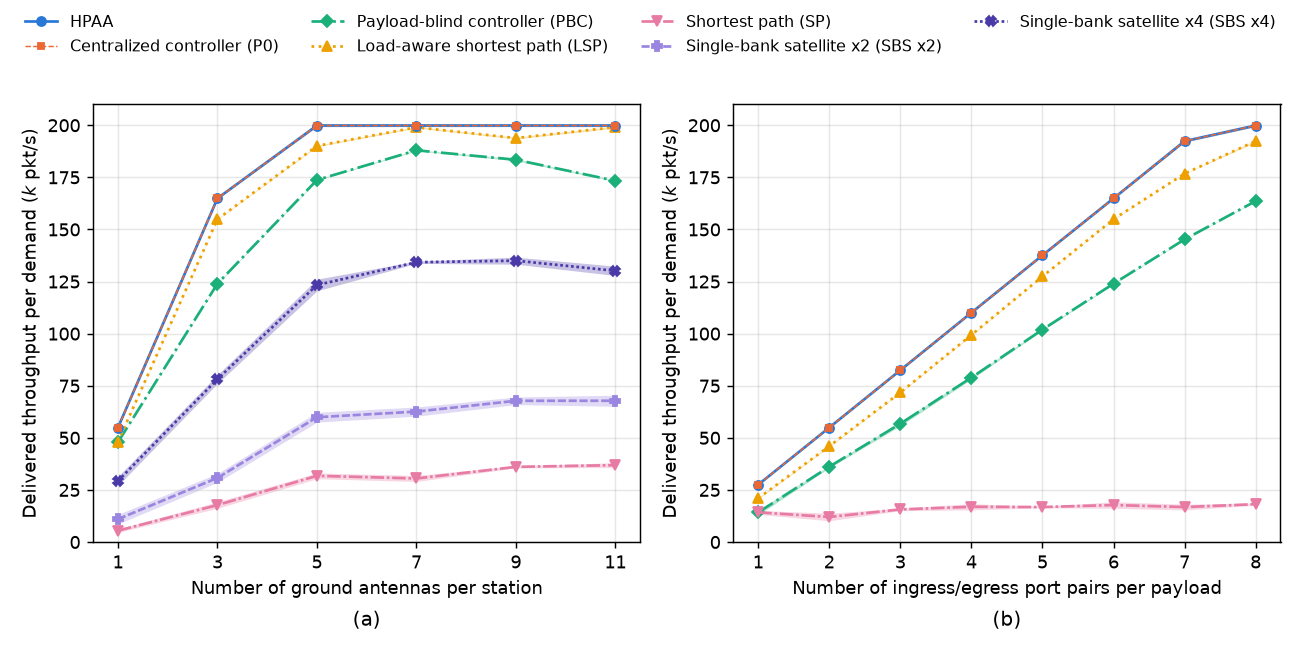}
  \caption{Sensitivity Analysis of the Delivered Throughput.}
  \label{fig:sens}
\end{figure*}

We assess the robustness of HPAA with a fixed mean requested rate of
$200$~kpkt/s per demand, while varying the number of antennas per ground
station as depicted in Fig.~\ref{fig:sens}a and the number of ingress/egress port pairs per
payload as depicted in Fig.~\ref{fig:sens}b. In both cases, HPAA matches
\textbf{(P0)}. HPAA reaches the full demand from $5$ antennas, and its
throughput grows by $27.5$~kpkt/s per port pair. Among the baselines,
LSP benefits most from additional antennas and ports, since its
load-aware weights spread the traffic over the added links. However, PBC
and SBS remain limited by overloaded payloads and single modem banks,
respectively. 
%SP yields almost constant throughput with the number of port pairs, since it loads a single modem bank per link.
 % SP remains far below the other schemes in both Fig.~\ref{fig:sens}a
 % and Fig.~\ref{fig:sens}b: since each demand follows a single shortest path and, with pinned ports, enters every satellite through a single ingress bank, its rate is capped by one modem bank ($\mu = 100$~kpkt/s), shared by all the demands crossing the same link. In Fig.~\ref{fig:sens}a, additional antennas spread the demands over more access satellites, raising SP from $5.5$ to about $32$~kpkt/s at $5$ antennas before it saturates below $37$~kpkt/s. In Fig.~\ref{fig:sens}a, additional port pairs never split the traffic of a link, so SP stays almost flat at $12$--$18$~kpkt/s.
SP performs worst because it is bottlenecked by its single-path, pinned-port structure, capping link rate at one modem bank. %($\mu = 100$~kpkt/s). 

\subsubsection{Computation Time}
\begin{figure}[H]
  \centering
  \includegraphics[scale=0.5]{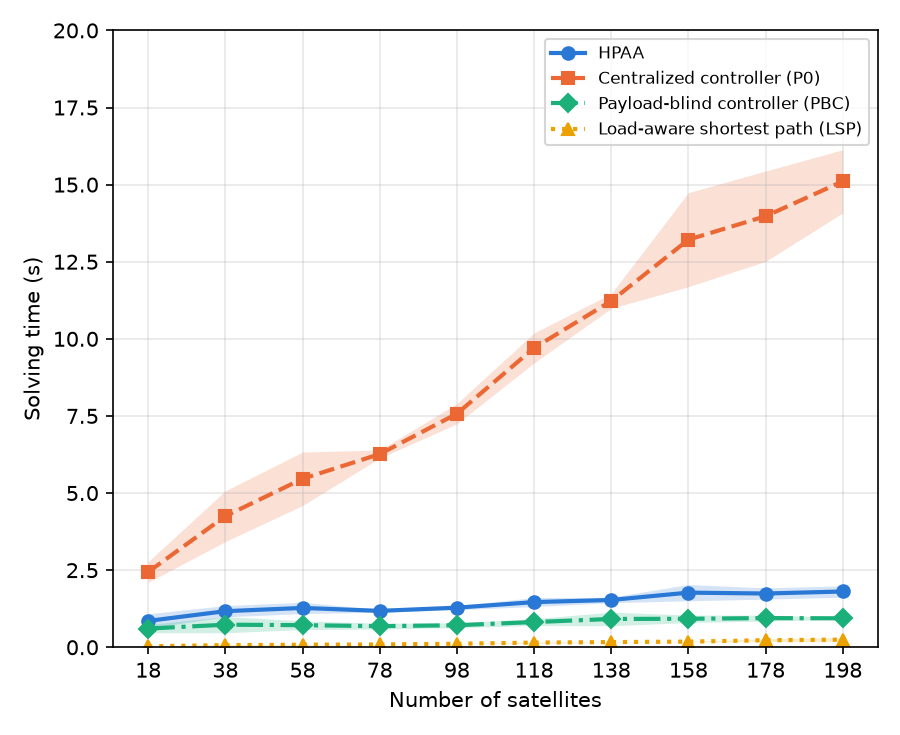}
  \caption{Computation Time  vs.\ Number of Satellites.}
  \label{fig:cost}
\end{figure}

We study the solving time per snapshot as a
function of the number of satellites.
%, averaged over $5$ snapshots per constellation size.  
Fig.~\ref{fig:cost} illustrates that HPAA is $8.3$ times faster than P0
with $198$ satellites, and its time grows slowly with the
constellation size, as each satellite's problem has a fixed size. In
contrast, the time of P0 grows steadily, since its single problem
includes the modem banks of every satellite. PBC and SBS are faster, but they ignore the payload constraints and thus deliver lower throughput
with substantial packet loss, as shown in Fig.~\ref{fig:sim_results}. 

%Overall, HPAA
% scales with the constellation size: it achieves the same allocation as
% the centralized solution while its solving time stays below $2$~s.

\section{Conclusion}\label{sec:conclusion}
This paper proposed a hierarchical SDN-based TE framework
for satellite constellations with regenerative payloads,
based on a two-level graph capturing both constellation-level
traffic allocation and intra-payload routing across interconnected
modem banks. Simulation results showed that the proposed HPAA
algorithm outperformed baseline control and routing strategies
in terms of delivered throughput, packet loss, and end-to-end
delay. A sensitivity analysis over the number of antennas per
ground station and the number of ingress and egress ports per
payload confirmed these gains. These improvements came at the
cost of higher computation time than conventional routing,
while remaining below that of the ideal centralized controller.
Future work will investigate multi-timescale TE for dynamic
traffic demands with heterogeneous requirements, as well as
deep reinforcement learning (DRL) approaches to optimize
long-term network performance.

% =========================================================================

%\IEEEtriggeratref{5}
\bibliographystyle{IEEEtran}
\bibliography{references}

%\newpage

\end{document}